\documentclass[conference]{IEEEtran}
\usepackage{cite}
\usepackage{amsmath,amssymb,amsfonts}
\usepackage{tikz}
\usepackage{algorithmic}
\usepackage{graphicx}
\usepackage{textcomp}
\usepackage{xcolor}
\def\BibTeX{{\rm B\kern-.05em{\sc i\kern-.025em b}\kern-.08em
    T\kern-.1667em\lower.7ex\hbox{E}\kern-.125emX}}
\begin{document}

\title{Decoding Imagined Speech using Wavelet Features and Deep Neural Networks 
}

\author{\IEEEauthorblockN{Jerrin Thomas Panachakel}
\IEEEauthorblockA{\textit{Indian Institute of Science} \\
Bangalore, India \\
jerrinp@iisc.ac.in}
\and
\IEEEauthorblockN{A.G. Ramakrishnan}
\IEEEauthorblockA{\textit{Indian Institute of Science} \\
Bangalore, India \\
agr@iisc.ac.in}
\and
\IEEEauthorblockN{T.V. Ananthapadmanabha}
\IEEEauthorblockA{\textit{Voice and Speech Systems} \\
Bangalore, India\\
tva.blr@gmail.com}}

\maketitle

\begin{abstract}
This paper proposes a novel approach that uses deep neural networks for classifying imagined speech, significantly increasing the classification accuracy. The proposed approach employs only the EEG channels over specific areas of the brain for classification, and derives distinct feature vectors from each of those channels. This gives us more data to train a classifier, enabling us to use deep learning approaches. Wavelet and temporal domain features are extracted from each channel. The final class label of each test trial is obtained by applying a majority voting on the classification results of the individual channels considered in the trial. This approach is used for classifying all the 11 prompts in the KaraOne dataset of imagined speech. The proposed architecture and the approach of treating the data have resulted in  an average classification accuracy of 57.15\%, which is an improvement of around 35\% over the state-of-the-art results.
\end{abstract}


\section{\label{sec:1} Introduction}

Speech is one of the most basic and natural form of human communication. However, nearly 70 million people have speech disabilities around the world. Speech disability due to complete paralysis prevents people from communicating with other through any modality.  It will greatly help such people, if by some means we are able to decode his/her thoughts, commonly referred to as ``imagined speech'' \cite{brigham2010imagined}.

The interest in imagined speech dates back to the days of Hans Berger, who invented electroencephalogram (EEG) as a tool for synthetic telepathy \cite{la1999history}. Although it is almost a century since the first EEG recording, the success in decoding imagined speech from EEG signals is rather limited. One of the major reasons for the same is the very low signal-to-noise ratio (SNR) of EEG signals.

The potential of the recent developments in the field of machine learning, such as deep neural networks (DNN) has not been exploited fully in the field of decoding imagined speech, since such techniques require a huge amount of training data. In this paper, we selected 11 EEG channels that cover the cortical areas involved in speech imagery. For each imagined word, each of the EEG channels so selected is considered as an independent input signal, thus providing more training data. This is in contrast to the earlier approaches concatenating the features to form a single feature vector.

Our new approach has been validated using the KaraOne dataset \cite{zhao2015classifying} and we have obtained accuracy values better than the state-of-the-art results reported in the literature for the same dataset.

The rest of the paper is organized as follows: Section \ref{sec:9} describes prior work in the literature in the field of decoding imagined speech. Section \ref{sec:3} describes the dataset and the procedure for generating the feature vectors. Section \ref{sec:4} describes the proposed DNN classifier in some detail. The results obtained are given in section \ref{sec:5}.

\section{\label{sec:9} Related Work in the Literature}
This section briefly describes the work in the field of decoding imagined speech, reported over the last decade.

C.S. DaSalla \textit{et al.} developed a brain-computer interface (BCI) system based on vowel imagery \cite{dasalla2009single} in the year 2009. The objective was to discriminate between the imagined vowels, \textit{/a/} and \textit{/u/}. The experimental paradigm consisted of three parts:
\begin{enumerate}
    \item Imagined mouth opening and imagined vocalisation of vowel \textit{/a/}.
    \item Imagined lip rounding and imagined vocalisation of vowel \textit{/u/}.
    \item Control state with no action.
\end{enumerate}
 
Using common spatial pattern (CSP) generated spatial filter vectors as features and nonlinear support vector machine (SVM) as the classifier, they achieved accuracies in the range of 56\% to 72\% for different subjects. As noted by Brigham \textit{et.al} \cite{brigham2010imagined}, the relatively higher accuracy obtained might have arisen because of the additional involvement of motor imagery.
 
Following a similar approach, Wang Li \textit{et al.} in 2013 developed a system to distinguish between two monosyllabic Chinese characters meaning ``left'' and ``one'' \cite{wang2013analysis}. Visual cue was provided to the subject to instruct him/her on the character to be imagined. When the cue disappeared, the subject had to repeatedly imagine the character in his/her mind as many times as possible for a duration of 4 sec. They obtained an accuracy of around 67\%.

In 2010, Brigham\textit{ et al.} came up with an algorithm based on autoregressive (AR) coefficients and k-nearest neighbor (k-NN) algorithm for classifying two imagined syllables \textit{/ba/} and \textit{/ku/} \cite{brigham2010imagined}. In this experiment, the subjects were given an auditory cue on the syllable to be imagined, followed by a series of click sounds. After the last click, the subjects were instructed to imagine the syllable once every 1.5 sec for a period of 6 sec. They reported an accuracy of around 61\%.

In 2016, Min \textit{et.al} used statistical features such as mean, variance, standard deviation, and skewness for pairwise classification of vowels (\textit{/a/}, \textit{/e/}, \textit{/i/}, \textit{/o/}, and \textit{/u/}) using extreme learning machine (ELM) with radial basis function. In their experimental paradigm, auditory cue  was provided at the beginning of the trial to inform the subject as to which  vowel was to be imagined. After the auditory cue, two beeps were played, after which the subject had to imagine the vowel heard during the beginning of the trial. An average accuracy of about 72\% was reported.

In 2017, Nguyen, Karavas and Artemiadis \cite{nguyen2017inferring} came up with an approach based on Riemannian manifold features for classifying four different sets of prompts:

\begin{enumerate}
\item Vowels (\textit{/a/}, \textit{/i/} and \textit{/u/}).
\item Short words (``\textit{in}'' and ``\textit{out}'').
\item Long words (``\textit{cooperate}'' and ``\textit{independent}'').
\item Short-long words  (``\textit{in}'' and ``\textit{cooperate}'').
\end{enumerate}
The accuracy reported for the four sets of prompts are  49.2\%, 50.1\%, 66.2\% and 80.1\%, respectively. This dataset is one amongst the few imagined speech datasets that are available in the public domain and is referred to as the \textit{``ASU dataset''}. More information about this dataset is given in Section \ref{sec:2}.

Balaji \textit{et al.} in 2017  investigated the use of bilingual imaginary speech, namely English ``\textit{Yes''} \& ``\textit{No}''	and Hindi ``\textit{Haan}'' (meaning ``\textit{yes}'') \& ``\textit{Na}'' (meaning ``\textit{no}'') for an imagined speech based BCI system \cite{balaji2017eeg}. Principal component analysis (PCA) was used for data reduction and an artificial neural network (ANN) was used as the classifier. Two specific sets of EEG channels corresponding to language comprehension and decision making were utilized. An interesting part of the experimental protocol was that there was no auditory or visual cue and the subjects were instructed to imagine the answer to a binary question posed either in English or Hindi. The study reported an accuracy of 75.4\% for the combined English-Hindi task and quite a surprisingly high accuracy of 85.2\% for classifying the decision.

In 2017, Sereshkeh \textit{et al.} came up with an algorithm based on features extracted using discrete wavelet transform (DWT) and regularized neural networks for classifying the imagined decisions of ``\textit{yes}'' and ``\textit{no}'' \cite{sereshkeh2017eeg}, similar to the work by Balaji \textit{et al.} They reported an accuracy of about 67\%.

In 2018, Cooney \textit{et al.} \cite{cooney2018mel} used Mel frequency cepstral coefficients  (MFCC) as features and SVM as classifier to classify all the 11 prompts in the KARAONE dataset \cite{zhao2015classifying}. The prompts consisted of seven phonemic/syllabic prompts (\textit{/iy/}, \textit{ /uw/},  \textit{/piy/}, \textit{/tiy/}, \textit{/diy/}, \textit{/m/}, \textit{/n/}) and four words (``pat'', ``pot'', ``knew'' and ``gnaw''). A maximum accuracy of only 33.33\% was achieved. The lower accuracy might have arisen because of a larger number of choices, unlike the binary choice employed in the previous works.

In a recent work \cite{jerrin-aibec}, Jerrin et al. used deep neural networks (DNN) for the first time to classify imagined speech. The specific task was to classify imagined words ``in'' and ``cooperate''. The features used were based on discrete wavelet transform and a DNN with three hidden layers was employed. The highest accuracy reported was around 86\%.

{\def\arraystretch{2}\tabcolsep=10pt
	\begin{table*}[h!]
	\centering
	\caption{Comparison of the mean cross-validation accuracies in percentage obtained using different methods (given in each column) in classifying 11 imagined prompts in the KaraOne dataset. ``s01'' to ``s08'' are the participant IDs. }
\begin{tabular}{c|c|c|c|}
\cline{2-4}
\multicolumn{1}{l|}{}                  & \multicolumn{3}{c|}{\textbf{Method}}                                     \\ \hline
\multicolumn{1}{|c|}{\textbf{Subject}} & \textbf{SVM+MFCC \cite{cooney2018mel}} & \textbf{DT+MFCC \cite{cooney2018mel}} & \textbf{\begin{tabular}[c]{@{}c@{}}Proposed \\ Method (DNN)\end{tabular}} \\ \hline
\multicolumn{1}{|c|}{s01}              & 22.27             & 24.52            & 43.02                                                                     \\ \hline
\multicolumn{1}{|c|}{s02}              & 33.33             & 31.06            & 60.91                                                                     \\ \hline
\multicolumn{1}{|c|}{s03}              & 23.62             & 15.12            & 84.23                                                                     \\ \hline
\multicolumn{1}{|c|}{s04}              & 15.31             & 21.14            & 45.78                                                                     \\ \hline
\multicolumn{1}{|c|}{s05}              & 14.84             & 11.41            & 37.43                                                                     \\ \hline
\multicolumn{1}{|c|}{s06}              & 20.86             & 21.17            & 60.81                                                                     \\ \hline
\multicolumn{1}{|c|}{s07}              & 26.08             & 26.84            & 75.07                                                                     \\ \hline
\multicolumn{1}{|c|}{s08}              & 23.15             & 18.37            & 49.98                                                                     \\ \hline
\multicolumn{1}{|c|}{Average:}         & 22.43             & 21.20            & 57.15                                                                     \\ \hline
\end{tabular}
\label{t1}
	\end{table*}

\section{Dataset Used for the Study and Methods\label{sec:3}}
\subsection{\label{sec:2} The KaraOne Dataset}
The KaraOne dataset \cite{zhao2015classifying} has been used for our study. The KaraOne dataset consists of EEG data captured during the imagination and articulation of 11 prompts, which included 7 phonemic/syllabic prompts (iy, uw, piy, tiy, diy, m, n) and 4 words derived from Kent's list of phonetically-similar pairs (i.e., pat, pot, knew, and gnaw). The data was captured at 1 KHz sampling rate using SynAmps RT amplifier. The electrode placement was based on the 10/10 system \cite{chatrian1985ten}. 

Each data recording trial had four stages:
\begin{enumerate}
    \item A 5-second rest state.
    \item A stimulus state in which an auditory and a visual cue were provided to the participant.
    \item A 5 seconds imagined speech state.
    \item An articulation state.
\end{enumerate}

We followed the same preprocessing steps as in \cite{zhao2015classifying}, which included ocular artifact removal using blind source separation \cite{gomez2006automatic}, band-pass filtering from 1 to 50 Hz and a Laplacian spatial filtering.

\subsection{Wavelet Feature Extraction}
	
In our work, instead of concatenating the features obtained from several channels, each channel is treated as a distinct input. This is possible because of the high correlation present between the signals of various channels \cite{ramakrishnan2016reconstruction}. The following 11 channels only have been chosen to be used in our work, based on the involvement of the underlying brain regions in the production of speech \cite{marslen2007morphology,alderson2015brain}:
\begin{enumerate}
    \item `C4': postcentral gyrus
    \item `FC3': premotor cortex
    \item `FC1': premotor cortex
    \item `F5': inferior frontal gyrus, Broca's area
    \item `C3': postcentral gyrus
    \item `F7': Broca's area
    \item `FT7': inferior temporal gyrus
    \item `CZ': postcentral gyrus
    \item `P3': superior parietal lobule
    \item `T7': middle temporal gyrus, secondary auditory cortex
    \item `C5': Wernicke's area, primary auditory cortex
\end{enumerate}

This choice of channels is also backed by the common spatial patterns (CSP)	analysis on imagined speech v/s rest state EEG data \cite{nguyen2017inferring}.
	
Since each EEG channel is considered as an independent data vector, algorithms that extract a single feature vector from the entire set of EEG channels (such as Reimannian manifold features used by Nguyen \textit{et.al} \cite{nguyen2017inferring} and fuzzy entropy features \cite{raghu2018novel}) cannot be used with the proposed architecture. 

For each trial, only the first 3000 samples (3 seconds) of collected data have been used for feature extraction. For extracting the temporal features, we divided the first 3000 samples into 4 equal blocks and extracted the following statistical features for each block:
\begin{enumerate}
    \item Root-mean-square
    \item Variance
    \item Kurtosis
    \item Skewness
    \item 3rd order moment
\end{enumerate}

Daubechies-4 (db4) wavelet is extensively used to extract features from EEG signals \cite{nicolas2012brain}. The 3 second-EEG signals are decomposed into 7 levels using db4 wavelet, for extracting the wavelet domain features. The above mentioned statistical features are extracted from the last approximation coefficients and for each of the last three detailed coefficients. This is performed to capture specific frequency bands that possess information on the cortical activity corresponding to the speech imagery. Hence, there are 20 temporal domain features and another 20 wavelet domain features adding up to feature vectors of dimension 40. 

\section{Details of the DNN Classifier\label{sec:4}}
A DNN with two hidden layers is used as the primary classifier. Since the dimension of the feature vector is 40, the number of neurons in the input layer is also 40.  Each dense hidden layer consists of 40 neurons. Also, dropout and batch normalization layers are added after each dense layer. The dropout ratio is 10\% for the two hidden layers. The activation function of all the layers except the first hidden later is the rectified linear unit. The activation function of the first hidden layer is hyperbolic tangent. The activation function of the  output layer is \textit{softmax}. Loss function is \textit{categorical cross-entropy}. This DNN architecture is adopted based on cross-validation performance of several DNN architectures. 

Because of the availability of very limited data, only cross-validation is performed. Since we have derived 11 feature vectors (one per each chosen channel) per trial, 11 outputs are obtained for each trial, one each for each channel. The final decision for each trial is then based on majority or hard voting of the 11 outputs.

\section{Results and Comparison with the Literature\label{sec:5}}
	
Five-fold cross-validation is performed on the pre-processed data of each participant. During cross-validation, it is ensured that all the channels corresponding to a trial are either in the training set or in the test set. This is important, since the presence of a couple of channels from the test trials in the training set can lead to high spurious accuracy due to data leakage. The cross-validation results obtained are listed in Table \ref{t1}, along with other results reported in the literature.

\section{\label{sec:l} Conclusion}

The present work shows that it is possible to treat each EEG channel as an independent data vector in order to increase the size of the training set for the purpose of decoding imagined speech using deep learning techniques. The proposed method gives around 35\% improvement in accuracy on an average over the state-of-the-art results.

\section{\label{sec:l} Acknowledgement}

The authors thank Dr. Frank Rudzicz, University of Toronto, Mr. Ciaran Cooney, Ulster University, Dr. Kanishka Sharma, Mr. Pradeep Kumar G. and Ms. Ritika Jain, Indian Institute of Science for the support extended to this work.

\bibliographystyle{IEEEtran}
\bibliography{mile3}
\end{document}